\documentclass[12pt]{article}
\usepackage[margin=0.7in]{geometry}
\usepackage{float}
\usepackage{amsmath}
\usepackage{amssymb}
\usepackage{graphicx}
\usepackage{cite}
\usepackage{color}
\usepackage{dcolumn}
\usepackage{bm}
\usepackage{booktabs}
\usepackage{newfloat}
\DeclareFloatingEnvironment[name={SI Figure}]{suppfigure}
\DeclareFloatingEnvironment[name={SI equation}]{suppequation}
\DeclareFloatingEnvironment[name={SI Table}]{supptable}
\usepackage{hyperref}
\begin{document}

\begin{flushleft}
{\LARGE
\textbf{Unveiling the Multi-fractal Structure of Complex Networks}
}
\\
Sarika Jalan$^{1,2,\ast}$, Alok Yadav$^{1}$, Camellia Sarkar$^{2}$ \& Stefano Boccaletti$^{3,4}$
\\ 
\it ${^1}$ Complex Systems Lab, Discipline of Physics, Indian Institute of Technology Indore, Khandwa Road, Simrol, Indore 453552, India\\
\it ${^2}$ Centre for Biosciences and Biomedical Engineering, Indian Institute of Technology Indore, Khandwa Road, Simrol, Indore 453552, India\\
\it ${^3}$ CNR-Institute of Complex Systems, Via Madonna del Piano, 10, 50019 Sesto Fiorentino, Florence, Italy\\
\it ${^4}$ The Italian Embassy in Israel, 25 Hamered st., 68125 Tel Aviv, Israel\\

$\ast$ Corresponding author e-mail: sarikajalan9@gmail.com
\end{flushleft}

\begin{abstract}
The fractal nature of graphs has traditionally been investigated by using the network's nodes as the basic units. Here, instead, we propose to concentrate on the graph's edges, and introduce a practical and computationally not demanding method for revealing changes in the fractal behavior of networks, and particularly for allowing distinction between mono-fractal, quasi mono-fractal, and multi-fractal structures. We show that degree homogeneity plays a crucial role in determining the fractal nature of the underlying network, and report on six different protein-protein interaction networks along with their corresponding random networks. Our analysis allows to identify varying levels of complexity in the species.
\end{abstract}

The fractal geometry of nature was first described by Mandelbrot already in 1967 \cite{BBM1975}, and the approach has been extensively used, since then, to gain insights into the underlying scaling of a variety of visually complex structures, like fracture surfaces of metals \cite{BBM1984}, strange attractors \cite{HGEH1983, EG1994}, diffusion \cite{BBM1974} and medical imaging  \cite{MIA2009} patterns, galaxies \cite{WLR1999}, and atomic spectra \cite{SM1996}, just to quote a few examples.  In complex networks, so far the focus has been investigating self-similarity \cite{SH2005} and fractal structure of skeleton \cite{JSK2007}, as well as growth models which capture the observed fractal behavior\cite{SJ2002,TV2010}.

A fundamental issue is identifying whether a system is a mono- or a multi-fractal, i.e. whether or not a unique fractal scaling spans all the different system's parts, regions or components. Multi-fractal analysis requires considering a physical measure: the number of nodes within a box of size, say $l$, has been used so far to analyze how the distribution of such number of nodes scales in a network, as the box size increases \cite{JSK2007}.
In this Letter, instead, we quarter the focus on the scaling of the number of edges in a partition-box of the network's adjacency matrix $\cal{A}$. Precisely, we consider a {\it spatial distribution} of the network's edges (instead of the network's nodes), this way making edges (the entries of the $\cal{A}$'s plane) as the basic units for the evaluation of fractal dimensions.

The aim is introducing a practical and computationally non demanding procedure, able to distinguish mono- from multi-fractal characters in complex networks. To do so, the distribution of 1's in square boxes of length $\epsilon$ in $\cal{A}$ is analyzed by the use of the box counting method (BCM) \cite{BCM1,BCM2,BCM3}, which considers only non overlapping boxes, thus preventing specific parts of the matrix from getting over-weighted due to systematic over-counting. Furthermore, the robustness of the results is pledged by reshuffling the nodes, i.e. by avoiding any form of biasness generated by a specific node indexing. Precisely, the supplementary material contains evidence that each shuffling step provides a
new ensemble of adjacency matrices \cite{supp}.

A schematic representation of the procedure is sketched in Fig. \ref{Fig1}. We start with partitioning the adjacency matrix into $\epsilon$-size boxes, and counting the number of boxes $n(\epsilon)$ with at least one non-zero entry (edge), with $\epsilon$ varying from 2 to $N/2$. $n(\epsilon)$ exhibits typically the scaling $n(\epsilon) \sim \epsilon^{D_{0}}$, with $D_0$ giving the dimension of the network. If $D_0$ is a non-integer number, the network is said to be {\it fractal}.
$D_0$, however, gives no information whatsoever on the system's  multi-fractality, which should be accounted for, instead, by a multi-fractal approach \cite{BCM1}. Let therefore digitalize $\cal{A}$ into $\epsilon$-size boxes, and let $n_i(\epsilon)$ be now the number of `1' entries in the $i^{th}$ element of the partition. The occurrence probability of 1's in the $i^{th}$ box of size $\epsilon$, denoted by $p_{i}(\epsilon)$, ranges therefore between $1/2N_c$ and $\epsilon^2/2N_c$. At each $\epsilon$ value, the $q^{th}$ moment (for all $-\infty < q < +\infty$ real numbers) of this probability is given by
\begin{figure}[t]
\centerline{\includegraphics[width=3.2cm,height=6.5cm,angle=90]{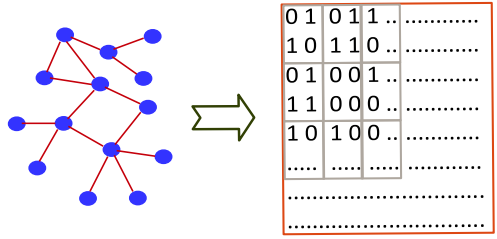}}
\caption{(Color Online). Schematic representation of the used box counting method. Left: original network. Right: partition of the adjacency matrix $\cal{A}$.}
\label{Fig1}
\end{figure}
\begin{equation}
P_{\epsilon}^q = \sum_{i=1}^{n(\epsilon)}[p_{i}(\epsilon)]^{q}.
\label{moment}
\end{equation}

The scaling exponent $\tau(q)$ is given by
\begin{equation}
\tau(q) = \lim_{\epsilon\to 0} \dfrac{ln P_{\epsilon}^q}{ln \epsilon},
\label{tauq}
\end{equation}
and is obtained from the slope of $P_{\epsilon}^q$ {\it vs.} $\epsilon$ at all $q$-value. Notice that $\tau(q)$ is here calculated {\it always} as ensemble average over the $P_{\epsilon}^q$ values obtained by shuffling the node indices.
The $q^{th}$-moment dimension $D_q$ is then given by
\begin{equation}
D_q = \dfrac{\tau(q)}{q-1}.
\label{Dq}
\end{equation}

\begin{figure}[b]
\centerline{\includegraphics[width=12.5cm,height=5cm]{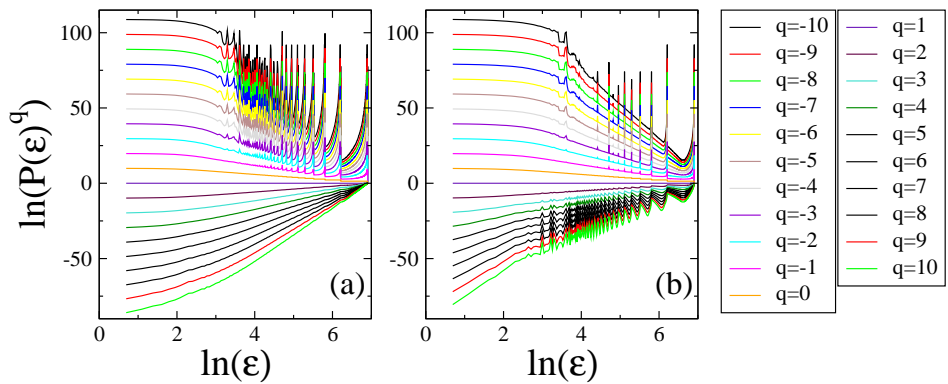}}
\caption{(Color Online). $P_{\epsilon}^q$ {\it vs.} $\varepsilon$ in a double logarithmic scale
with $q\in(-10,10)$ for (a) ER and (b) SF networks. $N=1,000$ and $\langle k \rangle=20$.}
\label{ER_SF_p_q}
\end{figure}

For each box of size $\epsilon$ and occupation probability $p_{i}(\epsilon)$, the singularity strength $\alpha_{i}$ is given by $p_{i}(\epsilon) = \epsilon^{\alpha_{i}}$, and at every $q$, $\alpha$ is evaluated as ${1}/{n(\epsilon)} \sum_{i=1}^{n(\epsilon)}\alpha_{i}$.
The singularity spectrum, $f(\alpha)$, is related with $\tau(q)$ by means of the Legendre transform
\begin{equation}
f(\alpha) = q\alpha - \tau(q),
\label{spectra}
\end{equation}
with $\alpha = {d\tau(q)}/{dq}$ being the H$\ddot{o}$lder exponent, and $f(\alpha)$ indicating the dimension of the subset scaling with $\alpha$.

A multi-fractal structure is indicated by the following marks \cite{BCM1}: {\it i)} multiple slopes of $\tau(q)$ {\it vs.} $q$; {\it ii)} a non constant value of $D_q$ {\it vs.} $q$,  and {\it iii)} $f(\alpha)$ {\it vs.} $\alpha$ covering a broad range instead of being accumulated at nearby non-integer values of $\alpha$.

\begin{figure}[t]
\centerline{\includegraphics[width=15cm,height=5cm]{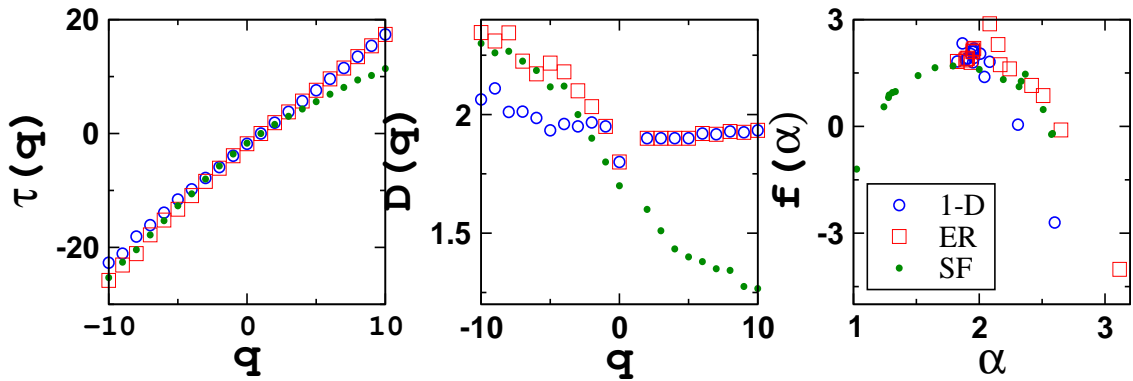}}
\caption{(Color Online). $\tau(q)$ (a) and $D_q$ (b) {\it vs.} $q$. (c) $f(\alpha)$ {\it vs.} $\alpha$. In all plots results are reported for 1-d lattices, ER and SF networks with $N=1,000$ and $\langle k \rangle=10$. All quantities are calculated for 50 random realizations of the networks generated by reshuffling indices 100,000 times for each realization. The left and right region slopes of $\tau(q)$ for the 1-d lattices, ER and SF networks are, respectively, 2.1 and 1.9; 2.4 and 1.9; 2.3 and 1.2.}
\label{1D_ER_SF_prop}
\end{figure}

We start by briefly discussing the effect of different reshuffling of the nodes' labeling, in particular (i) random reshuffling, and (ii) degree-based reshuffling. For an $\epsilon$-size box (located at the $(i,j)$ coordinate in the adjacency matrix plane and denoted as $b_{ij}$)
the probability of occurrence of 1's is $p_{\epsilon}^{b_{ij}} \propto \langle k \rangle$ in case i) (detailed derivations are provided in Supplementary Material), and is therefore independent of the network architecture.
Case ii) leads instead to interesting behavior, as the probability of occurrence of 1's
depends on the heterogeneity present in the networks. For 1-d lattices, where no degree
heterogeneity is present, random and degree-based reshuffling provide the same results \cite{supp}. At variance, Erd\"{o}s-R\'{e}nyi (ER) and scale-free (SF) networks, which display varying levels of degree heterogeneity, exhibit different behaviors for the two cases. For the degree-based reshuffling one has $p_{\epsilon}^{b_{ij}} \propto \frac{\epsilon^2}{N^4} \frac{k^2}{\rho(k)^2}$,
where $k$ is the degree of a node inside the box $b_{ij}$ and $\rho(k)$ is the probability of occurrence of $k$-degree nodes in the network. Note that $p_{\epsilon}^{b_{ij}}$ depends upon the degree as well as the probability density of degree of all the nodes inside the box \cite{supp}. In the following, we first sort the nodes of a network in decreasing order of their degrees and assign their indices. As the usual case is that many nodes have the same degree, in order to avoid the possibility of the preference being conferred on a particular node in a set of the same degree nodes, we carry out several realizations of the shuffling of indices among the nodes with the same degree.

We now present the results for 1-d lattices, ER and SF networks, and real world protein-protein interaction (PPI) networks of six different species. With a range of $\epsilon$ values spanning from $2$ to
$N/2$, $P_{\epsilon}^q$ is evaluated for $q$ values ranging between $-10$ to $+10$. The slope of $P_{\epsilon}^q$ (calculated using Eq.~\ref{moment}) versus $\epsilon$ (on a doubly logarithmic scale) provides the estimation of the value of the scaling exponent at each $q$, denoted as $\tau(q)$ (Eq.~\ref{tauq}). For all networks, the value of the scaling exponent is zero at $q=1$ and non-zero at other $q$ values (see Fig.~\ref{ER_SF_p_q}). The $q$ range can be divided, then into the left ($q \leq 0$) and the right
($q > 1$) region. A same value of the slope of $\tau(q)$ in both regions
indicates a mono-fractal nature of the network, while different values are a signature of a multi-fractal structure. For a regular
lattice, all  nodes have the same degree. This
kind of uniformity results in a single value of the scaling exponent on both left and right regions (see Fig.~\ref{1D_ER_SF_prop}(a)).
While the slopes of $\tau(q)$ are indicative measures of the multi-,
mono- or non-fractality of the system, the range of $D_q$ values  gives the dimension(s) of
the corresponding networks. For 1-d lattices, the $D_q$ values shrink to a very narrow range about $2$, confirming
again the mono-fractal character of the graph, while the singularity spectrum, $f_{\alpha}$ (calculated using Eq.~\ref{spectra}) shows that, for a certain range of $\alpha$, most of the $f_{\alpha}$ values are accumulated in a nearby non-integer region,  with only a few points deviating from this behavior.

The case of ER networks is far different. There, by construction, a new edge does not have preference to connect with specific nodes \cite{Newman_book}, leading to approximately similar degrees of all the nodes, and to an adjacency matrix where the 1 entries are dispersed in the entire plane, without clusters or patterns. The associated homogeneity gives nearly close values of scaling exponents on the the left and right regions on performing multi-fractal analysis (see Fig.~\ref{1D_ER_SF_prop}(a) and consult the Supplementary Material for the analysis of the impact of degree homogeneity on multi-fractal dimensions). Further, on plotting $D_q$ {\it vs.} $q$, one finds find that the right region exhibits exactly the same behavior as that of the 1-d lattice, whereas the left region deviates significantly from a mono-fractal pattern (see Fig.~\ref{1D_ER_SF_prop}(b)). The conclusion is  that ER networks can be said to exhibit a multi-fractal nature in the left region of the spectrum, and a quasi-mono-fractal character for $q > 1$.
When SF networks are taken into consideration, the growth mechanism producing them has a bias towards high degree nodes getting more and more edges (as preferential attachment induces new nodes to get linked to existing
high degree nodes \cite{Newman_book}), and gives a large spread in the degrees of the network's units.
In the adjacency matrix, this is reflected by few rows and columns (corresponding to high degree nodes) with a very large number of entries, with the majority of rows and columns having only very few 1's. Two large structures in the adjacency matrix then arise: one is a strip structure with a dense accumulation of 1's, and another is the sparse dispersion of entries in rest of the space. The
mixing of these two patterns yields different scaling exponents, and is indicative of a multi-fractal behavior (see Fig.~\ref{1D_ER_SF_prop}(a)).

The probability of occurrence of 1's in the $i^{th}$ box of size $\epsilon$, $p_{i}(\epsilon)$, ranges between $p_{min}(\epsilon)=1/2N_c$ and $p_{max}(\epsilon)=\epsilon^2/2N_c$. On considering the $q^{th}$ moments of these probabilities (Eq.~\ref{moment}), one finds that in the positive $q$ region the $p_{max}^{q}(\epsilon)$ values dominate over the $p_{min}^{q}(\epsilon)$, while in the negative $q$ region the $p_{min}^{q}(\epsilon)$ values exhibit dominance over $p_{max}^{q}(\epsilon)$ values with increasing magnitude of $q$. This means that the behavior of the densely packed boxes of entries in the adjacency matrix plane are reflected in the right $q$ region, while the sparsely packed boxes are determining the behavior of the left $q$ region. The conclusion is that the relevant character of the system (in terms of high density of edges) is described by the right region, while generally small  fluctuations in the system's fractal properties are accounted for the left region (as very small changes in the $q^{th}$ moments of the
probabilities get magnified in the negative $q$ region).

In 1-d lattices, all nodes have exactly the same degree, thus rendering uniform the distribution of 1's across the adjacency matrix planes. As a consequence, there are only two probabilities of occurrence of 1's: for all the boxes in the off-diagonal region, the probability is $\frac{2N_{c}\epsilon^{2}}{(N^{2}-N)}$, while for all the boxes lying in the diagonal region, the probability is
$\frac{2N_{c}(\epsilon^{2}-\epsilon)}{(N^{2}-N)}$. The slight kink in the $D_q$ values in the left region can be therefore attributed to the probabilities corresponding to the diagonal regions.

\begin{figure}[t]
\centerline{\includegraphics[width=14cm,height=10cm]{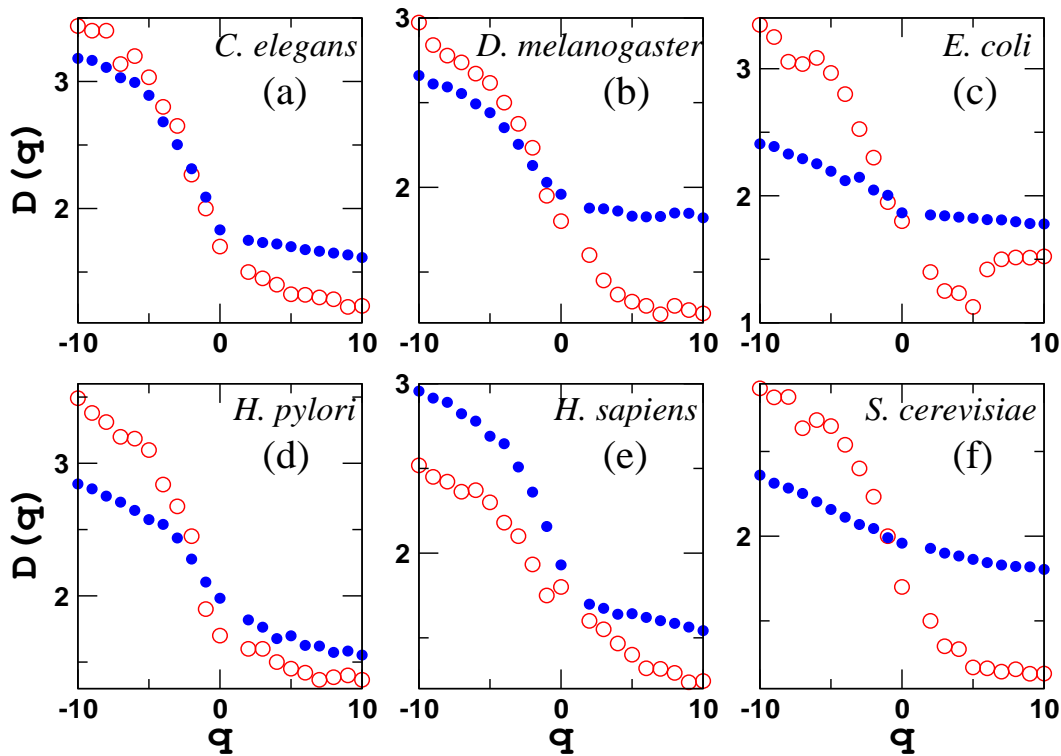}}
\caption{(Color online). $D_q$ {\it vs.} $q$ for PPI networks (open circles) of
six species, along with their corresponding ER random networks (closed circles) having
the same size and average degree. The results for each of the corresponding ER random networks are presented for 100,000 node reshufflings, and  for 50 realizations.}
\label{PPI_dq_vs_q}
\end{figure}

The degree distribution of a ER random network follows a Poisson statistics. Though most of the nodes have degree close to $\langle k \rangle$, there are nodes having higher and lower degrees as well.  The probability of occurrence of 1's in a box depends on the degree distribution, and deviates from that of a 1-d lattice. The deviation $\delta$ in the homogeneity of node degrees is reflected in the lower and upper bounds for the probability of occurrence of 1's: for all the boxes in the off-diagonal region, the probability  is $(\frac{2N_{c}}{(N^{2}-N)} \pm \delta)\epsilon^{2}$, whereas for those in the diagonal region, the
probability is $(\frac{2N_{c}}{(N^{2}-N)} \pm \delta)(\epsilon^{2}-\epsilon)$. Though the distribution of 1's in the adjacency matrix plane of the ER random networks appears very similar to that of a 1-d lattice (se Supplementary Material), the deviation in homogeneity of node degrees
results in significant deviations within the left region of $D_q$, as small changes are actually magnified in this region..

For SF networks, two types of dominant structures (patterns) exist in the adjacency matrix plane. The first structure is made up of densely packed rows and columns of 1's, while the second structure is formed by sparse population of 1's. The mixing of these two structures give rise to wide variety of probabilities $p_{i}(\epsilon)$'s. For positive (negative) $q$ values, the principal contribution to the multi-fractal nature of the graph comes from the first (the second) structure. Thus, for negative $q$ values, SF networks exhibit the same behavior as ER random networks, while their behavior strongly differs from that of ER graphs for positive $q$ values.

Finally, we apply our measure for shedding light on the structure and nature of associations in real world protein-protein interaction (PPI) networks. Namely, we consider the PPI networks of six different species: the {\it C. elegans}, the {\it D. melanogaster}, the {\it E. coli}, the {\it H. pylori}, the {\it H. sapiens} and the {\it S. cerevisiae}. In order to draw a fair comparison among the different PPI networks, we construct corresponding ER random networks having the same size and average degree as the real networks, and perform a comparative multi-fractal analysis for 100,000 rounds of node reshufflings, and 50 realizations.
On plotting $D_q$ {\it vs.} $q$ for the PPI networks and their corresponding random controls, one finds that for {\it E. coli} and {\it S. cerevisiae} both left and right regions deviate significantly from their random counterparts (see Fig.~\ref{PPI_dq_vs_q}(c) and (f)), indicating a genuine multi-fractal behavior of the underlying networks, which is quite expected as all these species are known to display scale-free properties in the degree distribution \cite{SJ_2014PhysicaA}. In the cases of  {\it D. melanogaster} and {\it H. pylori}, the extent of deviation of the PPI networks from their corresponding ER graphs is less evident (see Fig.~\ref{PPI_dq_vs_q}(b) and (d)). Interestingly, for {\it C. elegans}, the PPI network very closely resembles the behaviour of its corresponding random network (see Fig.~\ref{PPI_dq_vs_q}(a)).

In the case of {\it H. sapiens}, the behavior is very different as compared to all other PPI networks. Indeed, in contrast to the other five graphs, the {\it H. sapiens} PPI network exhibits a fractal behavior intermediate between ER random networks and 1-d lattices in the left region (see Fig.~\ref{PPI_dq_vs_q}(e) and \ref{1D_ER_SF_prop}(b)), indicating some kind of structural homogeneity in the underlying system, as also revealed recently through an analysis based on random matrix theory \cite{SJ_2014PhysicaA}.  The multi-fractal analysis supports and corroborates the hypotheses that real world biological systems are formed based on different designing principles, which are then reflected in difference in their dimensionality.

In conclusion, this Letter provides a practical and computationally not demanding method for inspecting the complexity of real world systems, by analyzing the scaling behaviour of edges in the underlying networks. Our study reveals that networks of the
same size and average degree can have completely different fractal character and nature,  depending upon how edges
are distributed in the networks. Further, we demonstrated how homogeneity in node degrees influences the scaling behavior of the underlying networks, and a deviation from degree homogeneity leads to significant changes in the graph's dimensionality. Our results point out the importance of edge distribution in determining the behavior of the system, and have diverse applications. For instance, different biological networks such as protein-protein interaction networks, metabolic networks, transcription regulatory networks, neural networks (which are based on different basic design and functional principles) may display similar fractal behavior depending upon similarity in their edge distributions. It is indeed important to remark that the underlying biological systems have evolved and survived over a long span of time without any major changes in their properties \cite{Cell_bio}. Additional examples demonstrating the importance of edge distribution are the edge-server network which facilitates the search on the internet by Google \cite{internet, internet1}, and the identification of influential nodes in social networks which takes into account the arrangement of the edges in the graph \cite{social}. Therefore the easy method for multi-fractal analysis presented here can be of interest and use for a better understanding and description of complexity in these systems. It could be furthermore interesting to investigate multifractality in networks constructed based on a given ordering of nodes, which would result in an adjacency matrix with a more pronounced block structure (specifically more dense diagonal blocks). Looking at the differences between a random permutation of the nodes and the ordering of the nodes that gives the best community structure can be one of the dimensions of future investigation.

SJ and AY acknowledge the Department of Science and Technology grant 25(0205)/12/EMR-II and the grant CSIR 09/1022(0013)/2014-EMR-I. SJ thanks J. N. Bandyopadhyay for useful discussions in the initial
phase of the work.

\newpage

\cleardoublepage
\centerline{{\Large \bf Supplementary Material}}
\vspace{1cm}
\section{Random reshuffling and degree-based reshuffling of the nodes}
\vspace{0.1cm}

In our work, two different methods of reshuffling the nodes have been considered.

\begin{enumerate}
\item The labeling of the nodes is reshuffled randomly.
Usually saturation of measurements occurs for a number of reshuffling being 10 times larger with respect to the number of connections.
\item A degree based reshuffling, where nodes are first sequenced based on their degree, and reshuffling operations
are made only among those nodes which have the same degree. For this case, the saturation of measurements is reached
much earlier.
\end{enumerate}

\subsection{Random reshuffling}
For random reshuffling, the probability of occurrence of 1's in boxes of the adjacency matrix plane is
proportional to $\langle k \rangle$. In the following, we analytically derive the differences.

For a network of size $N$, upon shuffling randomly the nodes' indices, a node of  degree $d_i$ (say)
can have an index $j$ (say) without any relation between $i$ and $j$.
Therefore, the probability for any node (say $u_i$) to have index $j$ is given by $p_{(u_{i})(j)} = 1/N$.
Let us now represent each box by its position in the adjacency matrix plane.
$b_{ij}$ represents a box which starts from $(i,j)$ in the adjacency matrix plane. 
\begin{suppfigure}
\centerline{\includegraphics[width=0.7\columnwidth]{SM_Fig1.eps}}
\caption{The effect of random reshuffling and degree-based reshuffling on the adjacency
matrix. (Left) ER random networks, (middle) SF networks
and (right) 1-d lattices. In each panel, each dot is the position of an entry in the adjacency matrix plane.}
\label{Fig1}
\end{suppfigure}

The number of 1's in a box $b_{ij}$ is given as
\begin{equation}
x_{ij} = \sum_{l_{1}=i+1}^{i+\epsilon}\sum_{l_{2}=j+1}^{j+\epsilon}a_{l_{1}l_{2}},
\end{equation}
where $l_{1}$ and $l_{2}$ are the adjacency matrix coordinates inside a box.

$a_{l_{1}l_{2}}$=1, if $l_1$ is connected to $l_2$ and 0 otherwise, meaning that the contribution of $l_1$ row to $x_{ij}$
is proportional to the degree of $l_1$-index node, as the $l_2$-column is randomly allotted for
a fixed $l_1$. This condition is arising due to the fact
that for a fixed $l_1$, the probability that
$l_2$ is connected to $l_1$ node is proportional to the degree of $l_1$ node.

Therefore,the  probability of any $a_{l_{1},i}$$ \neq 0 \, \, \, \forall \, \, \, i$ is given as
\begin{equation}
p_{a_{l_{1},i}}=\frac{1}{N}d_{l_1},
\end{equation}

and the probability of any $a_{l_{1},i}$$\neq 0 \, \, \forall \, \, \, i$ in
a box is
\begin{equation}
p_{a_{l_{1},i}}^{b_{ij}}=\frac{1}{N}d_{l_1}\epsilon.
\end{equation}

$x_{ij}$ is given by
\begin{equation}
x_{ij}=\sum_{l_{1}=i+1}^{i+\epsilon}\sum_{l_{2}=j+1}^{j+\epsilon}a_{ij}\neq0
\nonumber
\end{equation}

Let $p_{\epsilon}^{b_{ij}}$ be the probability that a given box $b_{ij}$ of size $\epsilon$ has at least one entry $1$.
Then one has $p_{\epsilon}^{b_{ij}}=\sum_{l_1}p_{a_{l_{1},i}}^{b_{ij}}$ x Probability of $d_{l_1}$-degree node in the box of size $\epsilon$.
One then has

\begin{equation}
p_{\epsilon}^{b_{ij}} = \frac{\epsilon}{N} \sum_{l_1} d_{l_1} \frac{\epsilon}{N} = \frac{\epsilon^2}{N^2} \sum_{l_1} d_{l_1}
\nonumber
\end{equation}

\begin{equation}
d_{l_1} = \sum_{j=1}^{N} P_{l_1}(d_j).d_{j}
\nonumber
\end{equation}
where $P_{l_1}(d_j)$ is the probability of $d_{j}$ node being allotted $l_1$ index.

In random reshuffling method, all the nodes in a network have equal probability of occurring at a particular index. Hence, $P_{l_1}(d_j) = \frac{1}{N}$. Therefore,
\begin{equation}
d_{l_1} = \frac{1}{N}\sum_{j=1}^{N}d_{j} = \frac{\langle k \rangle N}{N} = \langle k \rangle
\nonumber
\end{equation}

Since $l_1$ index runs within a box having size $\epsilon$, hence
\begin{equation}
\sum_{l_1} d_{l_1} = \epsilon \langle k \rangle
\label{sum_l1}
\end{equation}

Therefore,
\begin{equation}
p_{\epsilon}^{b_{ij}} = \frac{\epsilon^{3}\langle k \rangle}{N^2}
\end{equation}
This indicates that on considering random reshuffling of the nodes, the probability of 
occurrence of 1's in a box of size $\epsilon$ is directly proportional to the average degree 
of the underlying network. Further, to capture minor changes in this probability, 
with respect to various degree distributions, we consider 1-d, ER and SF networks.
\begin{suppfigure}[t]
\centerline{\includegraphics[width=0.7\columnwidth]{SM_Fig2.eps}}
\caption{Various multi-fractal measures plotted for the case of random reshuffling.}
\end{suppfigure}

In the 1-d lattices with circular boundary condition, all the nodes have exactly
the same degree which is equal to the average degree of the network. Thus each box 
exhibits exactly the same behaviour. For the ER random networks, the degree of nodes are 
closely distributed about the average degree of the underlying network. Therefore,
probability of occurrence of 1's in each box has small deviations from the average behaviour. 
In the case of SF networks, a very high degree heterogeneity is present in the networks, 
which leads to significant local deviations from the average behaviour, depending on the type of the node (high degree node or low degree node) occupying the boxes (see Fig.~S1 and S2).


To further assess the importance of the reshuffling method on the probability of 
occurrence of 1's in a box, in the following we adopt a different method of reshuffling 
considering degree homogeneity.

\subsection{Degree-based reshuffling}
In the degree-based reshuffling method, we arrange the nodes in a decreasing order of 
their degrees and assign them indices. To avoid any preferential allocation of indices 
between the same degree nodes, we reshuffle the indices between the same degree nodes. 
For a network of size $N$,
$\sum \rho(k)=1$ and $\sum N\rho(k)=N$, where $\rho(k)$ is the probability of occurrence of 
$k$-degree node and $N\rho(k)$ is number of nodes with degree $k$. Given that 
$0 \leqslant N\rho(k) \leqslant N$, 
\begin{itemize}
\item If $N\rho(k)=1$, any $k$-degree node having index $i$ is given as $i = \sum_{k=k_{max}}^{k}N\rho(k)$.

\item If $N\rho(k)>1$, any $k$-degree node having index $i$ is given as \\

\begin{equation}
i \in {\sum_{k=k_{max}}^{k+1}N\rho(k)+1, \sum_{k=k_{max}}^{k}N\rho(k)}
\label{index_range}
\end{equation}
\end{itemize}
For instance, let us consider a degree sequence viz. [200, 136, 80, 60, 60, 60, 30, 30, 22, 22, 22, 19 ......]. We want to find the range of indices which can have a node with degree
60.  From Eq.~\ref{index_range}, the upper and lower bound of indices that can have a node with
degree 60 is given as $(\sum_{k=200}^{80}N\rho(k)+1, \sum_{k=200}^{60}N\rho(k)) = ((1+1+1)+1, (1+1+1+3)) = (4, 6)$.

\begin{suppfigure}[h]
\centerline{\includegraphics[width=0.4\columnwidth]{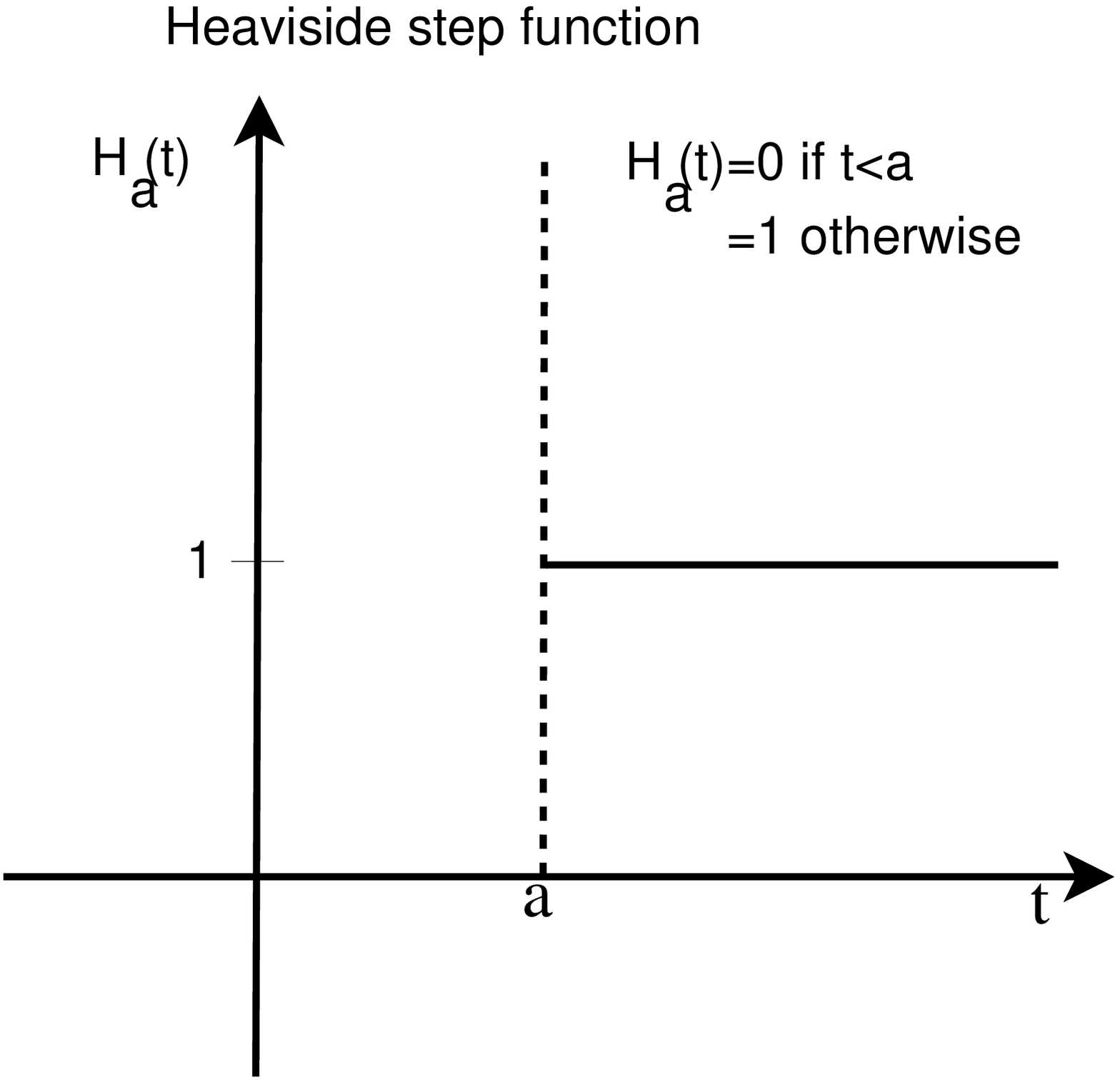}}
\caption{Graph representation of Heaviside step function used in Eq.~\ref{heaviside}.} 
\label{k-dist}
\end{suppfigure}

Since in the degree-based reshuffling, the nodes having similar degrees get reshuffled, 
the indices accessible to a particular degree (say $k_x$) are given by the probability 
\begin{equation}
p_{i}^{k_x}=\frac{H_{[\sum_{k=k_{max}}^{k_{x}+1}N\rho(k)]}(i) - H_{[\sum_{k=k_{max}}^{k_x}N\rho(k)]}(i)}{N\rho(k_x)},
\label{heaviside}
\end{equation} 
where $H_{a}(i)$ is the Heaviside step function (Fig.~S3).

Assuming there is no preference of the nodes being connected based on their degrees, 
we say that probability that 
any given $k$-degree node to be connected to any other 
node is equal to $\frac{k}{N}$.

Therefore, a box of $\epsilon$ size ($b_{ij}$) starting from a point in the adjacency matrix
plane ($i,j$) will have $x_{ij}$ number of 1's, given as;
\begin{equation}
x_{ij} = \sum_{l_{1},l_{2}=i+1,j+1}^{i+\epsilon,j+\epsilon}a_{l_{1}l_{2}},
\end{equation}
where $a_{l_{1}l_{2}}=1$, if $l_{1}$ and $l_{2}$ are connected and $0$, otherwise.

The probability of any $i^{th}$ node being connected to the $j^{th}$ node is given by

\begin{equation}
\begin{split}
p_{(a_{l_{1}l_{2}}=1)} &= \sum_{k_{1}k_{2}}p_{l_{1}}^{k_{1}}p_{l_{2}}^{k_{2}} \frac{k_{1}}{N} \frac{k_{2}}{N} \\
&= \sum_{k_{1}k_{2}}\frac{k_{1}k_{2}}{N^2} \frac{H_{[\sum_{k=k_{max}}^{k_{1}+1}N\rho(k)]}(l_{1}) - H_{[\sum_{k=k_{max}}^{k_{1}}N\rho(k)]}(l_{1})}{N\rho(k_{1})} \frac{H_{[\sum_{k=k_{max}}^{k_{2}+1}N\rho(k)]}(l_{2}) - H_{[\sum_{k=k_{max}}^{k_{2}}N\rho(k)]}(l_{2})}{N\rho(k_{2})}\\
&= \sum_{k_{1}k_{2}}\frac{k_{1}k_{2}}{N^{4}\rho(k_{1})\rho(k_{2})} (H_{[\sum_{k=k_{max}}^{k_{1}+1}N\rho(k)]}(l_{1}) - H_{[\sum_{k=k_{max}}^{k_{1}}N\rho(k)]}(l_{1}))(H_{[\sum_{k=k_{max}}^{k_{2}+1}N\rho(k)]}(l_{2})\\
&- H_{[\sum_{k=k_{max}}^{k_{2}}N\rho(k)]}(l_{2}))
\end{split}
\label{p_al1l2_heaviside}
\end{equation}

The probability of getting 1's in $b_{ij}$ box of $\epsilon$-size is given by
\begin{equation}
p_{(b_{ij})}^{\epsilon} = \sum_{l_{1}=i+1}^{i+\epsilon} \sum_{l_{2}=j+1}^{j+\epsilon} p_{(a_{l_{1}l_{2}}=1)}
\label{p_bij_heaviside}
\end{equation}

Now, according to the degree sequence of the network and the position $(i,j)$, 
$p_{(b_{ij})}^{\epsilon}$ can assume various values which are discussed in the following 
subsections.

\subsubsection{\it Case 1: The nodes have only one type of degree (say $k$) in the 
box $b_{ij}$.}  The indices of $\epsilon$-size box, i.e. 
($i+1$ to $i+\epsilon$, $j+1$ to $j+\epsilon$), which lies within the range given by 
Eq.~\ref{index_range} can have only $k$-degree nodes. Therefore, the probability of occurrence 
of $k$-degree node within the box is $\frac{1}{N\rho(k)}$, while the probability of 
occurrence of other degree node is zero. The probability of having 1's at 
$(l_{1},l_{2})$ position within the given box is drawn from Eq.~\ref{p_al1l2_heaviside} as 
\begin{equation}
p_{(a_{l_{1}l_{2}}=1)} = \frac{k^2}{N^4(\rho(k))^2}
\end{equation}

Since $p_{(a_{l_{1}l_{2}}=1)}$ is the same for all $a_{l_{1}l_{2}}$ within the box, 
the consequence is that
\begin{equation}
p_{b_{ij}}^{\epsilon} = \epsilon^{2} p_{(a_{l_{1}l_{2}}=1)} = \frac{k^{2}\epsilon^{2}}{N^{4}\rho(k)^{2}}
\label{eq14}
\end{equation}

\subsubsection{\it Case 2: The nodes have two types of degrees (say $k$ and $k+1$) in the $b_{ij}$ box.} 
The indices of a $\epsilon$-size box, i.e. ($i+1$ to $i+\epsilon$, $j+1$ to $j+\epsilon$) lies within the range given by Eq.~\ref{index_range} for $k$ and $k+1$ degree nodes. 
Therefore, the probability of any $k$-degree node allotted within the box is 
$p_{i}^{k}=\frac{H_{[\sum_{k=k_{max}}^{k+1}N\rho(k)]}(i) - H_{[\sum_{k=k_{max}}^{k}N\rho(k)]}(i)}{N\rho(k)}$ and the probability of any $k+1$-degree node allotted within the box 
is $p_{i}^{k+1}=\frac{H_{[\sum_{k=k_{max}}^{k+1+1}N\rho(k)]}(i) - H_{[\sum_{k=k_{max}}^{k+1}N\rho(k)]}(i)}{N\rho(k+1)}$. The probability of occurrence of any other degree nodes is zero. 
Therefore, the probability of any $a_{l_{1}l_{2}}$ being $`1'$ within the box $b_{ij}$ 
is given by
\begin{equation}
\begin{split}
p_{(a_{l_{1}l_{2}}=1)} &= \frac{(k+1)^2}{N^{4}\rho(k+1)^2} (H_{[\sum_{k=k_{max}}^{(k+1)+1}N\rho(k)]}(l_{1}) - H_{[\sum_{k=k_{max}}^{(k+1)}N\rho(k)]}(l_{1}))(H_{[\sum_{k=k_{max}}^{(k+1)+1}N\rho(k)]}(l_{2}) - H_{[\sum_{k=k_{max}}^{(k+1)}N\rho(k)]}(l_{2}))\\
&+ \frac{k^2}{N^{4}\rho(k)^2} (H_{[\sum_{k=k_{max}}^{(k)+1}N\rho(k)]}(l_{1}) - H_{[\sum_{k=k_{max}}^{(k)}N\rho(k)]}(l_{1}))(H_{[\sum_{k=k_{max}}^{(k)+1}N\rho(k)]}(l_{2}) - H_{[\sum_{k=k_{max}}^{(k)}N\rho(k)]}(l_{2})) \\
&+ \frac{k(k+1)}{N^{4}\rho(k)\rho(k+1)}
[(H_{[\sum_{k=k_{max}}^{(k)+1}N\rho(k)]}(l_{1}) - H_{[\sum_{k=k_{max}}^{(k)}N\rho(k)]}(l_{1}))(H_{[\sum_{k=k_{max}}^{(k+1)+1}N\rho(k)]}(l_{2}) - H_{[\sum_{k=k_{max}}^{(k+1)}N\rho(k)]}(l_{2})) \\
&+ (H_{[\sum_{k=k_{max}}^{(k+1)+1}N\rho(k)]}(l_{1}) - H_{[\sum_{k=k_{max}}^{(k+1)}N\rho(k)]}(l_{1}))(H_{[\sum_{k=k_{max}}^{(k)+1}N\rho(k)]}(l_{2}) - H_{[\sum_{k=k_{max}}^{(k)}N\rho(k)]}(l_{2}))].
\end{split}
\label{eq15}
\end{equation}

Therefore, the probability of having 1's inside the box is
\begin{equation}
\begin{split}
p_{\epsilon}^{b_{ij}} &= \sum_{l_{1}=i+1}^{i+\epsilon} \sum_{l_{2}=j+1}^{j+\epsilon} p_{(a_{l_{1}l_{2}}=1)} \\
&= \sum_{l_{1}=i+1}^{i+\epsilon_{1}} \sum_{l_{2}=j+1}^{j+\epsilon_{2}} p_{(a_{l_{1}l_{2}}=1)}^{(k+1)^2} + \sum_{l_{1}=i+1}^{i+\epsilon_{1}}
 \sum_{l_{2}=j+\epsilon_{2}+1}^{j+\epsilon} p_{(a_{l_{1}l_{2}}=1)}^{k(k+1)} \\
&+ \sum_{l_{1}=i+\epsilon_{1}+1}^{i+\epsilon} \sum_{l_{2}=j+1}^{j+\epsilon_{2}} p_{(a_{l_{1}l_{2}}=1)}^{k(k+1)} + \sum_{l_{1}=i+\epsilon_{1}+1}^{i+\epsilon} \sum_{l_{2}=j+\epsilon_{2}+1}^{j+\epsilon} p_{(a_{l_{1}l_{2}}=1)}^{k^2}.
\end{split}
\label{eq16}
\end{equation}

On substituting Eq.~\ref{eq15} into Eq.~\ref{eq16}, one has
\begin{equation}
\begin{split}
p_{\epsilon}^{b_{ij}} = \frac{1}{N^4} [\epsilon_{1} \epsilon_{2} (\frac{k+1}{\rho(k+1)} - \frac{k}{\rho(k)})^{2} + \frac{\epsilon^{2}k^{2}}{\rho(k)^{2}} + \frac{\epsilon(\epsilon_{1}+\epsilon_{2})k}{\rho(k)}(\frac{(k+1)}{\rho(k+1)} - \frac{k}{\rho(k)})].
\end{split}
\end{equation}
\\

In comparison to the random reshuffling, the degree-based reshuffling highlights 
following differences for both the case.  For the case of degree-based reshuffling, the probability of 
occurrence of 1's within a box depends on the position of the box as well as the 
coordinates within the box (for instance, ($l_1$,$l_2$) $\in$ {($i+1$,$j+1$) ........ ($i+\epsilon$,$j+\epsilon$)}). 
More importantly, the probability of occurrence of 1's in the 
degree-based reshuffling method depends on $\frac{k}{\rho(k)}$, i.e., network's degree distribution.  Whereas, in the case of random reshuffling, the probability of occurrence of 1's 
within a box depends only on the average degree of the underlying network. 
Therefore, the degree-based reshuffling leads to different behaviours for 1-d lattices, 
ER random networks and SF networks of the same average degree and size (see Fig.~S1).

After discussing impact of reshuffling on the probability of occurrence of $1's$ in 
a particular box, in the following we derive expressions for probability distribution
as well as $q^{th}$ moment of fractal dimension
for various different types of the edge distributions. 

\section{Edge distribution of different networks on adjacency matrix plane}
For the edge distribution of random networks and 1-d lattices, we present three simplified
cases: (a) a strictly homogeneous distribution, (b) an approximately homogeneous distribution and (c) a strictly linear distribution.

\subsection{Case A: Strictly Homogeneous distribution }


Any $\epsilon$-size box in the adjacency matrix plane will have $\frac{\langle k\rangle\epsilon^2}{N}$ number of 1's. Here, we consider large size network to avoid any boundary effect while assigning the boxes.
Therefore, the probability of occurrence of 1's in a $i^{th}$ box of size $\epsilon$ is given as \newline

$p_i(\epsilon) = \frac{\epsilon^2}{N^2}$.\newline

Summing up these probabilities gives \newline

 $P_{\epsilon}=\sum_{i=1}^{i=\frac{N^2}{\epsilon^2}}p_i(\epsilon)=\sum_{i=1}^{i=\frac{N^2}{\epsilon^2}}\frac{\epsilon^2}{N^2} = 1$.\newline

The $q^{th}$-moment of the probability of occurrence of 1's for a $\epsilon$ size box is given as \newline

$P_{\epsilon}^{q}=\sum_{i=1}^{i=\frac{N^2}{\epsilon^2}}[p_i(\epsilon)]^q=\sum_{i=1}^{i=\frac{N^2}{\epsilon^2}}(\frac{\epsilon^2}{N^2})^q = \frac{\epsilon^{2(q-1)}}{N^{2(q-1)}}$.\newline

By taking log of both the sides, one has \newline

$ln(P_{\epsilon}^{q})=2(q-1)ln(\epsilon)-2(q-1)ln(N)$\newline


This is an equation for a straight line with a slope $2(q-1)$, which we define as the 
scaling exponent $\tau(q)$.\newline


The $q^{th}$-moment dimension of the system $D(q)$ defined as $\frac{\tau(q)}{q-1}$ gives a 
single dimension i.e. 2. 

The singularity spectrum, $f(\alpha)$ is related with $\tau(q)$ by means of 
Legendre tranform\newline

$f(\alpha)= q\alpha -\tau(q)$.\newline

where $\alpha = \frac{d\tau(q)}{dq}$.

Substituting the value of $\tau(q)$, $\alpha = 2$ and $f(\alpha) = 2q - 2(q-1) = 2$.

This indicates that a homogeneous distribution over a large system (where boundary effect can be neglected) leads to a non-fractal structure with dimension 2.
\newline

\begin{suppfigure}
\centerline{\includegraphics[width=0.8\columnwidth]{SM_Fig4.eps}}
\caption{(Color Online) Comparison between, respectively, $ln(P_i^q(\epsilon))$ vs $ln(\epsilon)$ plot, $\tau(q)$ vs $q$ plot, $D(q)$ vs $q$ plot and $f(\alpha)$ vs $\alpha$ plots for strictly (top panel) and approximately (bottom panel) homogeneous distribution of edges.}
\label{Fig4}
\end{suppfigure}

\subsection{Case B : Approximately Homogeneous distribution }
In this case, we assume that the number of 1's in the $i^{th}$ box deviates from the 
homogeneous distribution of 1's by a small amount $r_i$ generated randomly for each box 
with alternately changing sign in such a way that sum over all the multi-fractal measures is 
normalized. This gives the following expression for the multi-fractal measure,\newline

$P_i(\epsilon)=\sum_{i=1}^{\frac{N^2}{\epsilon^2}}(\frac{\epsilon^2}{N^2}(1+(-1)^{i+1}r_i))=1$,\newline

which further gives us the following restriction on the random numbers ($r_i$) to be generated,\newline

$\sum_{i=1}^{\frac{N^2}{\epsilon^2}}(\frac{\epsilon^2}{N^2}(-1)^{i+1}r_i)=0$\newline

Now, $q^{th}$ moment of $P_i(\epsilon)$ can be written as,\newline

$P_i^q(\epsilon)=\sum_{i=1}^{\frac{N^2}{\epsilon^2}}(\frac{\epsilon^2}{N^2}(1+(-1)^{i+1}r_i))^q$\newline

For small $r_i$ values, the q-th moment can be approximated to,\newline

$P_i^q(\epsilon)=\sum_{i=1}^{\frac{N^2}{\epsilon^2}}(\frac{\epsilon}{N})^{2q}(1+(-1)^{i+1}qr_i)$\newline

$P_i^q(\epsilon)=(\frac{\epsilon}{N})^{2(q-1)} + \sum_{i=1}^{\frac{N^2}{\epsilon^2}}(\frac{\epsilon}{N})^{2q}(-1)^{i+1}qr_i)$\newline

$P_i^q(\epsilon)=(\frac{\epsilon}{N})^{2(q-1)} (1+ (\frac{q\epsilon^2}{N^2})\sum_{i=1}^{\frac{N^2}{\epsilon^2}}(-1)^{i+1}r_i)$\newline

Taking log on both the side, we get,\newline

$ln(P_i^q(\epsilon))=2(q-1)ln(\epsilon)-2(q-1)ln(N)+ ln(1+(\frac{q\epsilon^2}{N^2})\sum_{i=1}^{\frac{N^2}{\epsilon^2}}(-1)^{i+1}r_i)$\newline

By using $ln(1+x)$ approximation for small $x$, we get,

$ln(P_i^q(\epsilon))=2(q-1)ln(\epsilon)-2(q-1)ln(N)+ {(\frac{q\epsilon^2}{N^2})\sum_{i=1}^{\frac{N^2}{\epsilon^2}}(-1)^{i+1}r_i} -\frac{1}{2}((\frac{q\epsilon^2}{N^2})\sum_{i=1}^{\frac{N^2}{\epsilon^2}}(-1)^{i+1}r_i)^2 + \frac{1}{3}((\frac{q\epsilon^2}{N^2})\sum_{i=1}^{\frac{N^2}{\epsilon^2}}(-1)^{i+1}r_i)^3-...$\newline

Also, $\tau(q)= \lim_{\epsilon\rightarrow 0} \frac{ln(P_i^q(\epsilon)}{ln(\epsilon)}$ gives \newline

$\tau(q)= 2(q-1)$\newline

$D(q)=2$\newline

One can see that theoretically this approximation does not affect values of 
$\tau(q)$ and $D(q)$ from those of case A. However, numerical evaluation shows significant 
deviations of these values from those of the case A. For these two cases of 
strictly homogeneous and approximately homogeneous distribution, on plotting 
$ln(P_{i}^{q}(\epsilon))$ vs $ln(\epsilon)$ we find that the strictly homogeneous case 
has a straight line behaviour with existence of various slopes indicating
non-fractal structure with dimension 2. Whereas the approximately homogeneous case manifests 
a deviation leading to non-fractal dimensions distributed very closely about 2 which is
in fact a multi-fractal structure (see Fig.~S4). This indicates that one of the 
reasons behind $ln(P_{i}^{q}(\epsilon))$ vs $ln(\epsilon)$ plot deviating from 
the linear behaviour can be attributed to the non-homogeneity in the degree
distribution.

\section{$\tau_q$ versus $q$ for PPI Networks}
On plotting $\tau(q)$ versus $q$ for the PPI networks and their corresponding random 
controls, for {\it E. coli} and {\it S. cerevisiae}, we find a significant deviation 
exhibited by the PPI networks from their corresponding random controls (see Fig. S5(c) and (f)), indicating multi-fractal behaviour of the underlying networks. This
is quite expected owing to their scale-free nature of degree distribution [18]. 
The extent of deviation between the PPI networks of $D. melanogaster$ and $H. pylori$ and 
their corresponding ER random networks diminishes (Fig. S5(b) and (d)), 
while noticeably for $C. elegans$ the PPI network very closely resembles the 
behaviour of its corresponding random network (Fig. S4(a)). Interestingly, $H. sapiens$ 
exhibits a completely different behaviour from rest of the model organisms. The left region 
of $H. sapiens$ PPI network exhibits a slope which is lower than that of its corresponding ER random network (see Fig. S5(e)), thus indicating that though the underlying system maintains 
an universality in large scale properties with the other PPI networks, minor fluctuations in 
fractal behaviour captured by the left region indicates a homogeneity in the 
underlying structure. These features of the PPI networks are magnified and better understood through the behaviour of $D_q$ versus $q$, discussed in length in the paper.
\begin{suppfigure}[bh]
\centerline{\includegraphics[width=0.7\columnwidth]{SM_Fig5.eps}}
\caption{(Color Online) Plots of $\tau_q$ as a function of $q$ for PPI networks (circles) of
six species along with their corresponding ER random networks (triangles) having
the same size and average degree. The results for each of the corresponding ER random networks are presented for 100,000 node reshuffling done for 50 realizations.}
\label{PPI_tauq_vs_q}
\end{suppfigure}

\end{document}